\documentclass{ws-ijmpd}
\usepackage[super,compress]{cite}
\usepackage{graphicx}
\usepackage[numbers]{natbib}
\begin{document}
\markboth{Alexander F. Zakharov}{The black hole at the Galactic
Center: observations and models}

%
\catchline{}{}{}{}{}
%

\title{The black hole at the Galactic
Center: observations and models}

\author{Alexander F. Zakharov$^{1,2,3,4,5}$}

\address{$^{1}$ National Astronomical Observatories of Chinese Academy of Sciences,
          20A Datun Road,  100012 Beijing, China}

\address{$^{2}$ Institute of Theoretical and Experimental
Physics,
            B. Cheremushkinskaya 25,             117218 Moscow,  Russia\\
zakharov@itep.ru}

\address{$^{3}$Bogoliubov
Laboratory of Theoretical Physics, Joint Institute
  for Nuclear Research,
   141980 Dubna,
             Russia}

\address{$^{4}$National Research Nuclear University  MEPhI
(Moscow Engineering Physics Institute), 115409, Moscow,
 Russia}

\address{$^{5}$North Carolina Central University, Durham, NC
27707,
 USA}

\maketitle

\begin{history}
\received{Day Month Year}
\revised{Day Month Year}
\end{history}

\begin{abstract}

One of the most interesting astronomical objects is the Galactic
Center. It is a subject of intensive astronomical observations in
different spectral bands in recent years. We concentrate our
discussion on a theoretical analysis of observational data of bright
stars in the IR-band obtained with large telescopes. We also discuss
the importance of VLBI observations of bright structures which could
characterize the shadow at the Galactic Center. If we adopt general
relativity (GR) there are a number of theoretical models for the
Galactic Center, such as a cluster of neutron stars, boson stars,
neutrino balls, etc. Some of these models were rejected or the range
of their parameters is significantly constrained with consequent
observations and theoretical analysis. In recent years a number of
alternative theories of gravity has been proposed because there are
dark matter (DM) and dark energy (DE) problems. An alternative
theory of gravity may be considered as one  possible solution for
such problems. Some of these theories have black hole
solutions, while other theories have no such solutions. There are
attempts to describe the Galactic Center with alternative theories
of gravity and in this case one can constrain parameters of such
theories with observational data for the Galactic Center. In
particular, theories of massive gravity are intensively developing
and theorists have  overcome pathologies presented in initial
versions of these theories. In theories of massive gravity, a
graviton is massive in contrast with GR where a graviton is
massless. Now these theories are considered as an alternative to GR.
For example, the LIGO-Virgo collaboration obtained the graviton mass
constraint of about $1.2 \times 10^{-22}$~eV in their first
publication about the discovery of the first gravitational wave
detection event that resulted of the merger of two massive black
holes. Surprisingly, one could obtain a consistent and comparable
constraint of graviton mass at a level around  $m_{g} < 2.9 \times
10^{-21}$~eV from analysis of observational data on the trajectory
of the star S2 near the Galactic Center. Therefore, observations of
bright stars with existing and forthcoming telescopes such as the
European Extremely Large Telescope (E-ELT) and the Thirty Meter
Telescope (TMT) are extremely useful for investigating the structure
of the Galactic Center in the framework of GR, but these
observations also give a tool to confirm, rule out or constrain
alternative theories of gravity. As we noted earlier, VLBI
observations with current and forthcoming global networks (like the
Event Horizon Telescope) are used to check the hypothesis about the
presence of a supermassive black hole at the Galactic Center.

\end{abstract}

\ccode{PACS numbers: 04.20.-q, 04.30.Tv, 04.70.-s, 04.70.Bw,
04.80.Cc, 98.35.Jk}


\section{Centennial history of black holes in brief}

It is well-known that general relativity (GR) was discovered in
November 1915 \citep{Einstein_15,Hilbert_16}. German mathematician
K. Schwarzschild found a vacuum solution of GR equations
\citep{Schwarzschild_16}  (he submitted his article in January 1916
or only a few months after the creation of GR). Now we call the
Schwarzschild solution a spherically symmetric black hole. Initially
the Schwarzschild solution had no singularities because it described
only an external  Schwarzschild solution. In 1917, Hilbert generalized
the Schwarzschild solution \cite{Hilbert_17}, where he used other
coordinates which are called now Schwarzschild ones. The Hilbert
solution described both internal and external regions.
 There are
two types of singularities in the solution. The first type of
singularity is connected with a coordinate choice, this singularity
arises at the event horizon, and the second type of singularity is
connected with singularities in Riemannian tensor components at the
origin and the singularities correspond to infinite tidal forces and
these singularities cannot be removed with a coordinate choice.
Infinite physical quantities in the physical model indicate
pathologies or defects in such a theory. Usually, it is accepted
that a region near the origin should be described quantum gravity
but not with classical GR. A brief description of different aspects
of the Schwarzschild solution is presented in the interesting
article \cite{Eisenstaedt_82}. Many scientists (including A.
Einstein \citep{Einstein_39}) were rather skeptical in respect to
possible astronomical application of the Schwarzschild solution and
they thought that the Schwarzschild solution  cannot be realized in
nature  due to a presence of pathologies in the solution.
 In his textbook Einstein's assistant P. Bergmann expressed the same opinion  that the Schwarzschild solution is not realized in physical reality \cite{Bergmann_42}.
  Soon after the discovery of Fermi -- Dirac statistics, Soviet physicist J. Frenkel
 created a theory of white dwarfs \cite{Frenkel_28} and in the paper he discussed all basic properties of white dwarfs
 except an existence of maximal mass of such stars, but unfortunately this remarkable paper was unknown among astrophysicists \cite{Yakovlev_94}.\footnote{A comprehensive history of relativistic astrophysics development before WWII is given in \cite{Bonolis_17}.} A conclusion about an existence of maximal mass of white dwarfs was discussed in \cite{Anderson_29,Stoner_30,Stoner_31,Chandrasekhar_31,Landau_32,Chandrasekhar_34}.
  A connection of the Landau's result about an existence of maximal mass in white dwarfs \cite{Landau_32}
 with existence of maximal mass of neutron stars
  is discussed in \cite{Yakovlev_13} where it was noted that Landau's consideration can be applied successfully for a theory of neutron stars.
  However, when Landau obtained the result about maximal mass for white dwarfs he noted that the result is not trustworthy, therefore
  one should conclude there are regions in stars where laws of quantum mechanics and quantum statistics should be violated \cite{Landau_32}.
 \footnote{A. Eddington had the same opinion \cite{Eddington_35}.}
In 1939, Oppenheimer and Volkoff found the upper mass limit for
neutron stars \citep{Oppenheimer_39} while Oppenheimer and Snyder
showed there is the opportunity for black hole formation as a result
of stellar collapse \cite{Oppenheimer_39b}. At the beginning, the
result about the mass limit of neutron stars was looked very
controversial. Except for Landau's opinion that laws of quantum
mechanics should be violated in the interiors of stars, there were
two different opinions about the fate of massive stars. The first
opinion that there are stars with masses which are greater than the
mass limits for white dwarfs and neutron stars and an object which
is called now black hole should be formed\footnote{S. Chandrasekhar always
supported this point of view.}; the second opinion is that in the
process of their evolution, stars are losing mass and at the end of
their life stellar masses will be always smaller than the
Oppenheimer -- Volkoff limit.

For a distant observer, the collapse of a star into a black hole
lasts an infinite time, so the collapsing star  was earlier called a
frozen star or collapsing star \cite{Zeldovich_96}. The metric for a
collapsing star tends to be static or stationary (Schwarzschild or
Kerr) when a proper time for a distant observer tends to infinity.
In 1967, J. A. Wheeler suggested using the limiting (static or
stationary) metric for a physical model of a collapsing astronomical
object and he called it a black hole \citep{Wheeler_68}. The
proposed idea is very efficient since after a finite time, interval
differences between a dynamical metric and its limiting metric of
black hole started to be very small and it could be reasonable to
substitute a complicated dynamical metric with the simple static (or
stationary) case.

The Schwarzschild solution for a spherically symmetric black hole
has been known for more than century, but, there are only a few
cases where one really has to use the metric in the strong
gravitational field limit but not a weak one. However, even in a
weak gravitational field limit, one can expect deviations from the
Newtonian theory such as the famous GR effects: gravitational
redshift, relativistic deflection of light,  Mercury anomaly
(relativistic advance of apocenter) and Shapiro time delay.

In 1989, A. Fabian and co-authors considered possible shapes of
spectral lines emitted from a circular ring region near a black hole
\citep{Fabian_89} (see also \citep{Stella_90,Laor_91,Matt_93}). The
authors concluded the shapes of spectral lines could be a powerful
tool to evaluate a spin,  emission region for the line and a
position angle for a distant observer.

Based on observational data from  the Japanese ASCA X-ray satellite
for Seyfert  galaxy MCG-6-30-15, astronomers found signatures of the
relativistic iron $K\alpha$-line \citep{Tanaka_95}. The authors
concluded that an emission region is very close to the event
horizon, so that a spin parameter $a$ of the Kerr metric is very
close to 1, since the innermost stable circular orbits (ISCOs) in
the Kerr metric are located closer to the event horizon in
comparison with ISCOs in the Schwarzschild metric \citep{Tanaka_95}.
Therefore, one must use a Kerr metric in a strong gravitational
field limit to create adequate theoretical model to fit the
observational data. However, other authors
\citep{Karas_00,Turner_02,Dovciak_04,Murphy_09} expressed their
skepticism in respect to uniqueness of such a theoretical model.

Based on our previous results \cite{Zakharov_94,Zakharov_95}, we
write our own ray-shooting code to check conclusions about the
interpretation of observational data in the framework of the
theoretical model. We solved several relevant astrophysical problems
with our code
\cite{Zakharov_99,Zakharov_02,Zakharov_03b,Zakharov_03c,Zakharov_04,Zakharov_05R,Zakharov_06,Zakharov_04a,Zakharov_05,Zakharov_07,
Z_NCB_2007}, in particular, we generalize Matt et al.'s result
\cite{Matt_93} about the existence of extra peaks in relativistic
line shapes for the case of a Kerr black hole   if a distant
observer is located near the equatorial plane \cite{Zakharov_03a}
(initially Matt et al. \cite{Matt_93} established their result for a Schwarzschild
black hole case) while in \cite{ZKLR02} we showed that in principle
one could constrain a magnetic field, when  the magnetic field is
strong enough. One can find more recent reviews on theoretical
aspects of relativistic line shapes in
\cite{Fabian_10,Jovanovic_12}.

The Galactic Center (Sgr $A^*$) is one of the most attractive
objects for observations in different spectral bands from radio to
$\gamma$-ray (it is located at a distance around 8 kpc from our
Solar System). There are many theoretical models for the Galactic
Center including a dense cluster of stars \cite{Reid_09}, fermion
ball \cite{Munyaneza_02}, boson stars \cite{Jetzer_92,Torres_00} and
neutrino balls \cite{DePaolis_01},  however, the most popular and
natural model is a supermassive black hole. Different theoretical
models have different observational predictions, but sometimes
observational differences may be very small as it was shown for the
analyzed cases of shadow shapes in the framework of a boson star and
the conventional black hole model \cite{Vincent_16}.

The hypothesis about a presence of a supermassive black hole at the
Galactic Center has to be confirmed with observations. One could use
test bodies to evaluate gravitational potential at the Galactic
Center similar to Newton's procedure to derive the gravity law from
Kepler's laws or to E. Rutherford's analysis of $\alpha$-particle
trajectories to evaluate a potential and to understand the structure
of atoms.

Using VLBI observations in the mm-band with current and future world
wide networks such as the Event Horizon Telescope
\cite{Doeleman_09}, astronomers use photons to trace the black hole
metric at the Galactic Center. Researchers investigate a structure
of bright spots near the black hole to reconstruct the shadow shape.
To interpret such observational data, certainly astronomers have to
develop a theoretical model where a strong gravitational field plays
a key role \cite{Falcke_13,Johannsen_16}. Below we discuss aspects
of shadow formation in details.

Another option is to use bright stars or clouds of hot gas to trace gravitational potential at the Galactic Center.
 Astronomers still do not need a GR approach in a strong gravitational field
approximation to fit their observational data, but observational
facilities are developing so rapidly that in the future one should
use the GR approach in a strong gravitational field limit.  We
discuss results of observations of bright stars and conclusions from
these observations in the next section.

\section{Observations of the Galactic Center}

There are two groups of observers monitoring bright stars at the
Galactic Center. One group of observers uses the twin 10-meter
optical/infrared telescopes at Mauna Kea (Hawaii). Their aim is to
monitor stars in the IR band with a high angular resolution.  Andrea
Ghez is a professor at the University of California (Los Angeles)
and she leads this group. This group is a world leader in precise
measurements of trajectories of bright stars near the Galactic
Center. Some results of their observations and interpretations of
the results may be found in
\citep{Ghez_00,Ghez_03,Ghez_04,Ghez_05,Weinberg_05,Meyer_12,Morris_12}.

There is another group consisting of  European astronomers. They use
four  eight meter telescopes (VLT) at the Paranal Mountain. These
telescopes belong to European Southern Observatory. The group has
Reinhard Genzel as a leader. Important results of the group can be
found in the papers and references therein
\citep{Schodel_02,Genzel_03,Eckart_04,Eckart_05a,Eckart_05b,Eckart_06a,Eckart_06b,Eckart_07,Gillessen_09,Gillessen_12}.
Observational results of American and European groups are
complimentary and consistent.

Some time ago, the European group of astronomers claimed that they
observed the G2  gas cloud \citep{Gillessen_12} and noted that
similar to bright stars, the G2 gas cloud could be used as a good
tracer of gravitational potential at the Galactic Center. The
authors predicted that the cloud should be disrupted near its
pericenter\footnote{Some people use the word "peribothron" instead
of pericenter \cite{Frank_76}.} passage. However, subsequent
observations showed that the object was not disrupted after the
pericenter
 passage, therefore, it was concluded that
it is a dust-enshrouded star \citep{Phifer_13,Zajacek_14,Zajacek_15} or a young star \citep{Valencia_15}.

Recently, the European Southern Observatory  and Max Planck
Institute for Extraterrestrial Physics built  the Very Large
Telescope Interferometer which is called GRAVITY.  The
interferometer will be used for precise astrometrical observations
at a level around 10 micro-arcseconds
\citep{Eisenhauer_11,Blind_15}.  A science verification for GRAVITY
has been done in 2016, and the first scientific results were
published in 2017 \cite{GRAVITY_17}.

In the near future, the Thirty Meter Telescope (TMT) and European
Extremely Large Telescope (E-ELT) will be in action. Observations of
bright stars near the Galactic Center are listed as high priority
targets in scientific programs using these forthcoming facilities.

\section{Constraints on black hole parameters and gravity theories from trajectories of bright stars at the Galactic Center}
\subsection{Constraints on black hole parameters and extended mass
distribution}

As it was noted two groups of astronomers monitored bright stars at the Galactic Center for
decades,
see discussions of results of observations and their
interpretation in
\citep{Ghez_00,Ghez_03,Ghez_04,Ghez_05,Weinberg_05,Meyer_12,Morris_12,Schodel_02,Genzel_03,Eckart_04,Eckart_05a,Eckart_05b,Eckart_06a,Eckart_06b,Eckart_07,Gillessen_09,Gillessen_12,Zhang_15,Yu_16}.
Source 2 (S2 or S0-2) is a bright star with mass around 15~$M_\odot$.
It has a relatively high eccentricity  $e \approx 0.876$, the
orbital period $P \approx 15$ years and the pericenter distance is
around 120 AU. In 2008, astronomers finished observations for one
orbital period of this star. At the beginning of these observations,
astrometrical precision was at the level 10~mas, and since 2005 the
astrometrical precision has been better than 1~mas, which
practically coincides with the relativistic advance
\citep{Rubilar_01,NDIQZ_07,ZNDI_PRD_2007}.\footnote{That is the famous relativistic
effect or Mercury anomaly which was explained by A. Einstein in 1915
in the framework of GR.}

It is well-known that if there is a black hole alone (without a
stellar cluster and a concentration of dark matter near a black
hole) a test particle orbit is not elliptical, but there is
so-called relativistic advance. In the case of a Kerr black hole,
the relativistic advance depends on spin, but for the cases of S2
and S16, additional terms due to the presence of spin are roughly
100 times smaller than the classical value for the relativistic
advance \citep{Rubilar_01,NDIQZ_07,ZNDI_PRD_2007} (see also an
updated discussion in \cite{Saxton_16,Dokuchaev_15}).

As it was noted earlier for an observer on the Earth that the
relativistic advance per period is around 1 mas for star S2.
 However, if there is an extended mass distribution between
an apocenter and a pericenter, the distribution causes the
pericenter and apocenter to shift in direction, which is opposite to
the relativistic one. If only 5\% of black hole mass is distributed
as a bulk inside the S2 orbit, then an extended mass distribution
causes a pericenter shift in the opposite direction with respect to
the relativistic one. Therefore, in the future precise observations
of bright stars will help to support, rule out or constrain some
models of an extended mass distribution of stellar cluster and dark
matter. For instance, in paper \citep{ZNDI_PRD_2007} we ruled out
some theoretical models of dark matter distributions which were used
earlier by astroparticle physicists to explain $\gamma$-ray flux
from the Galactic Center with neutralino annihilation. Perhaps in
the future, astronomers will get so stringent constraints on
dark matter distribution that there will be no way to explain
$\gamma$-flux from the Galactic Center region with neutralino
annihilation \cite{Z_SPIG_08}.

\subsection{Constraints on  $R^n$ theory}

If we adopt conventional GR then we face the so-called dark matter
(DM) and dark energy (DE) problems. As it was discussed in
\cite{Zakharov_09}, a famous French astronomer U. J. Le Verrier
proposed a rather general approach to resolve some anomalies.
According to Le Verrier's proposal, the first option to resolve an
anomaly is to introduce an additional component in a model\footnote{Le
Verrier predicted the existence of Neptune with the approach.}, the
second option is to modify a fundamental law\footnote{Einstein used the
approach to explain the Mercury anomaly.}, the third option is a
clarification and improvement of a theoretical model. An
introduction of dark matter and dark energy within the standard
theory of gravity (GR) practically implements the first Le Verrier's
option.

Since there is a very slow progress in understanding and resolving the
puzzles of dark matter and dark energy, people have proposed
changing Einstein's gravity law to explain dark energy
($\Lambda$-term) as an entire gravitational effect (it is the second
Le Verrier's option). To change the gravity law theorists introduced
of a generalization of the classical Einstein -- Hilbert Lagrangian
and substitute a linear function of scalar curvature $R$ with
arbitrary function $f(R)$
\citep{Capozziello_02,Carroll_04,Capozziello_04,Capozziello_06a,Capozziello_06b,Capozziello_07,Capozziello_10,Capozziello_11}.
With such  kind of alternative of gravity one could explain many
cosmological phenomena, including the acceleration of the Universe
and rotation curves for spiral galaxies without the introduction of
dark matter. However,  these theories very often have no Newtonian
limit in a weak gravitational field approximation but  a validity of
Newton's gravity law had been checked for different length scales
and different astronomical objects, therefore, one has to  obtain
the Newtonian limit for gravity for such objects in the framework of
suggested theories of gravity. For instance, an alternative theory
of gravity introduced in paper \cite{Carroll_04} does not fit
observational data in our Solar system. Another class of theories,
introduced in \citep{Capozziello_06a,Capozziello_06b} with the
Lagrangian $f(R)=R^n$ (which  corresponds to GR in the case $n=1$),
perfectly describes acceleration of the universe and rotation curves
in spiral galaxies and for these cases parameter $n$ has to be
significantly different from 1 or more precisely $n \in [1.5, 3]$, however, as was
shown in our papers \citep{ZNDI_PRD_2006,MZDNI_NCB_2007} to fit
observational data for Solar system one needs $n\approx 1$,
therefore one comes to a contradiction.

In paper \citep{BJBZ_PRD_12} we used $R^n$ theory of gravity to fit
observational data for star S2 by analyzing observational data from
the Keck and VLT telescopes. We concluded that $n\approx 1$ and as
it was noted earlier, it contradicts a suitable range of $n$
obtained from cosmological SNeIa data and rotation curves of spiral
galaxies. In papers \citep{Zakharov_14ASR,Borka_15_NSP} we
considered $R^n$ theory of gravity to fit observational data for
star S2 in  the framework of $R^n$ theory
for the case when there is additionally a stellar cluster around the black
hole.
We found that the presence of a stellar cluster did not
significantly change our previous conclusions that $n\approx 1$ and
a gravity theory should be very close to GR.

\subsection{Constraints on Yukawa gravity theory}

Sometimes, $f(R)$ theories of gravity use the Yukawa limit in a weak
gravitational field approximation
\citep{Capozziello_07b,Capozziello_09b,Cardone_11,Napolitano_12} and the authors successfully
implemented Yukawa like gravity to fit observational data for
rotation curves of spiral galaxies and galactic clusters. Earlier,
assuming the existence of a dilaton it was discussed a presence of the Yukawa
gravitational interaction and its possible experimental manifestations \cite{Fujii_71}.

In paper \cite{BJBJZ_JCAP_13}, we used Yukawa gravity to constrain
parameters of the potential with star S2 orbit data obtained from
observations with the VLT and Keck telescopes, so we considered the
following potential
\begin{equation}
\Phi \left( r \right) = -\dfrac{GM}{(1+\delta)r}\left[ {1 + \delta
e^{- \left(\dfrac{r}{\Lambda} \right)}} \right], \label{equ01}
\end{equation}
\noindent where $\Lambda$ and $\delta$ are parameters of the Yukawa potential.

In paper \cite{BJBJZ_JCAP_13} we obtained the preferable range of
$\Lambda$ parameter in the case of star S2 and it has to be $\Lambda
\in [5000,  7000]$~AU and we found that it is very  hard to get the
constraints on  $\delta$ parameter due to degeneracies of the
$\chi^2$ function. We also found that $\delta=1/3$ could be used to
fit observational data for the orbit of S2. Earlier, the same value
has been used for galactic clusters in \citep{Capozziello_07b,Capozziello_09b} and
rotation curves of spiral galaxies \citep{Cardone_11}.

\subsection{Constraints on massive graviton theory}

In \cite{FP} Fierz and Pauli introduced a theory of massive gravity
(see a simple introduction in \citep{Visser_98}), however later a
lot pathologies have been found in  theories of massive gravity such
as the presence of ghosts and discontinuities
\citep{Zakharov_70,van_Dam_70}, however, in the last year theorists
overcame many technical problems
\citep{Rubakov_08,DeRham_12,DeRham_17} and now such a theory is
considered as an alternative for GR.
There are different ways to constrain graviton mass with observations \cite{Goldhaber_10,DeRham_17} and
we mention only a few opportunities to do it with experiments where the main goal is a detection of gravitational waves.

Many years ago it was proposed to use pulsar timing to detect gravitational waves
with long wavelengths \cite{Sazhin_78} and as it was shown in \cite{Lee_10,Lee_13} pulsar timing arrays could be used
to evaluate a graviton mass.

A simple expression for a Yukawa modification of a Newtonian potential could be written in the following form
\citep{Visser_98,Will_98,Will_14}
\begin{eqnarray}
  V(r) = \frac{GM}{r}
\exp(-r/\lambda_{\mathrm{g}}).
\end{eqnarray}
C. Will discussed an opportunity  to evaluate a graviton mass  with
an analysis of delays of gravitational waves with respect to
electromagnetic radiation from supernovae or GRBs. This idea has
been used by the LIGO-Virgo collaboration to evaluate a graviton
mass \cite{Will_98,Will_14}. Assuming that a graviton is massive,
one could use the well-known dispersion relation
\begin{equation}
E^2= \frac{m_g^2 c^4}{{1-(v_g^2/c^2)}}, \label{Dispersion_Eq}
\end{equation}
where $E$, $v_g$ and  $m_g$ are energy, velocity and mass of
graviton, respectively. Therefore, if a graviton is massive then
more energetic gravitons propagate faster than slower ones and
gravitational wave signal is different for theories with
 massive and massless gravitons. Second, as it was noted that if a graviton has a non-vanishing mass then there exists an additional time delay between electromagnetic and gravitational wave signals, so there is another opportunity to evaluate a graviton mass.

Based on analysis of S2 orbit data obtained with VLT and Keck
telescopes we found that at the 90\% confidence level we have
$\lambda_{\mathrm{g}} > 2900$~AU$ = 4.3 \times 10^{11}$ km or $m_{g}
< 2.9 \times 10^{-21}$~eV
\citep{ZMBJBJ_16,Zakharov_Quarks_16,ZJBBJ_MIFI_17,ZJBBJ_Baldin_17}.

In February  2016 the LIGO-Virgo collaboration published  a couple
of works where the authors reported the discovery of gravitational
waves from mergers of  black holes with masses of about $30~M_\odot$
\cite{Abbott_16}. Moreover, the authors claimed that they also found
the graviton mass constraint $1.2 \times 10^{-22}$~eV. Therefore,
the LIGO-Virgo collaboration not only discovered gravitational waves
and the existence of binary black holes with high stellar masses but
they also obtained a fundamental result about constraints on
possible generalizations of GR considering massive gravity theories.
   On  June 2, 2017 the LIGO-Virgo collaboration reported the discovery of the third GW event from merging of BHs with 31 and 19 solar masses at redshift z=0.19 and improved the graviton mass constraint $m_g < 7.7 \times 10^{-23}$~eV \cite{Abbott_17}. The LIGO graviton mass constraint is around 38
times better than our bound obtained from an analysis  of S2 star
trajectory, but our estimate is independent and it may be improved
with current and future observational data
\cite{Hees_PRL_17,Hees_17,Chu_17}. In October 2017 the LIGO collaboration
presented constraints on speed of gravity from analysis of times
of arrivals for gravitational wave and electromagnetic signals in
the merger of two neutron stars and kilonova explosion $-3\times
10^{-15} < (v_{g}-c)/c < 7 \times 10^{-16}$
\cite{Abbott_170817,Abbott_ApJL_170817} and one could derive a graviton
mass constraint from these inequalities and the dispersion
relation \cite{Zakharov_17_IHEP}.

\section{Shadows for the black hole at the Galactic Center}

In papers \cite{Falcke00,Falcke_00a,Melia01} the authors considered
a toy model where particles are accreting onto black hole and  emit
photons in different directions and these photons form a picture
which could be seen by a  distant observer. In this case people say
that  a shadow (a dark spot) around a black hole is formed (see also Figs. in \cite{Bardeen_73,chandra}).
If we ignore scattering then photons with
different energies propagate along the same geodesics.  For the case
of the Galactic Center, the size of the shadow is $D_S =3\sqrt{3}R_g
\approx 50~\mu$as where $R_g$ is an angular size of the Schwarzschild
radius for the black hole at the Galactic Center. The shadow may be
detectable at mm and sub-mm bands while in cm band scattering spoils
images of the black hole
\cite{Falcke00,Falcke_00a,Melia01,Melia_07}. In spite of the very
simple model, subsequent simulations and observations basically
confirmed these claims.

In recent years, reaching better angular resolution in mm-band for
the Sgr $A^*$ \citep{Shen_05,Doeleman_08} represented significant
progress, and Doeleman et al. found the smallest spot with a size
$37^{+16}_{-10}$ $\mu as$ \cite{Doeleman_08}. The observations have
been done with the VLBI technique. In the future, radio astronomers
plan to build the world wide VLBI network \cite{Doeleman_09} which
will be called the Event Horizon Telescope because the angular
resolution of the network will be better than the angular size of
the event horizon for the black hole at Galactic Center or the black
hole at the center of M87.

Based on ideas introduced in \cite{chandra, Holz02}, in the paper
\cite{ZNDI05} the authors considered different types of shadow
shapes for Kerr black holes and different position angles of a
distant observer. In this paper it is also found that for an equatorial
plane position of a distant observer, the maximal critical impact
parameter in the perpendicular direction to the equatorial plane
$|\beta_{\rm max}|=\sqrt{27}$ (in $GM/c^2$ units) and corresponding
impact parameter in perpendicular direction $\alpha_{\rm max}=2a$
(see details in \citep{ZNDI05}). The critical curve $\beta (\alpha)$
separates capture and scattering regions. It means that a shadow
size does not depend on spin, but a shadow shape depends on spin
(see corresponding Figs. in \cite{ZNDI05}).

We obtained analytical relations for shadows in Reissner -- Nordstr\"om and K\"ottler (or Schwarzschild -- de-Sitter) metrics and discuss observational consequences in papers
\citep{ZNDI05,ZNDI05b,ZDIN_NA_11,Z_PRD_14,Stuchlik_83,Zakharov_2014,Zakharov_JAA_15}.

A comprehensive review of observational signatures of supermassive black hole presence at the Galactic Center is given in \cite{Eckart_17},
where the authors discussed shadows not only for supermassive black hole but also for other theoretical models for the Galactic Center.

\section{Conclusions}

Observations of bright stars near the Galactic Center give us an unique opportunity
to investigate a presence of an extended mass distribution near the Galactic Center
and to check and constrain a suitable range of parameters for alternative theories of gravity.

More precise observations  will  come with the Event Horizon Telescope,
the GRAVITY interferometer or/and forthcoming large telescopes (E-ELT
and TMT) and probably many current claims will be clarified.

One can obtain the graviton mass constraint from an analysis of
S2 star trajectory and the bound is consistent and comparable with
the constraint presented recently by the LIGO collaboration.\cite{Zakharov_17_IHEP}

\subsection*{Acknowledgements}

The author thanks D. Borka, V. Borka Jovanovi\'c, F. De Paolis,  G.
Ingrosso, P. Jovanovi\'c, S. M. Kopeikin, Y. Lu,  A. A. Nucita, S.G.
Rubin, Z. Stuchl\'ik, B. Vlahovic for useful discussions. A. F. Z. thanks PIFI
grant 2017VMA0014 of Chinese Academy of Sciences, the Strategic
Priority Research Program of the Chinese Academy of Sciences (Grant
No. XDB23040100), NSF (HRD-0833184) and NASA (NNX09AV07A) at NASA
CADRE and NSF CREST Centers (NCCU, Durham, NC, USA) for a partial
support. A. F. Z. also thanks an anonymous referee for useful critical
remarks, Dr. James Wicker for critical remarks and his help to
improve the manuscript, and the organizers of the Third Conference
on Particle Physics and Astrophysics for their attention to the
contribution.


\end{document}